\magnification=1200

\hsize=125mm
\vsize=180mm
\hoffset=3mm
\voffset=12mm

\baselineskip=14pt

\font\twelvebf=cmbx12

\font\tenbb=msym10
\font\sevenbb=msym7
\font\fivebb=msym5
\newfam\bbfam
\textfont\bbfam=\tenbb \scriptfont\bbfam=\sevenbb
\scriptscriptfont\bbfam=\fivebb
\def\bb{\fam\bbfam}

\def\Cb{{\bb C}}
\def\Hb{{\bb H}}
\def\Rb{{\bb R}}
\def\Ab{{\bb A}}
\def\Bb{{\bb B}}

\def\Zb{{\bb Z}}

\def\Ac{{\cal A}}
\def\Hc{{\cal H}}

\def\Uc{{\cal U}}

\def\Diff{\mathop{\rm Diff}\nolimits}
\def\Aut{\mathop{\rm Aut}\nolimits}
\def\Trace{\mathop{\rm Trace}\nolimits}

\def\Tr{\mathop{\rm Tr}\nolimits}

\def\Gev{\mathop{\rm Gev}\nolimits}

\def\Int{\mathop{\rm Int}\nolimits}

\def\Lb{\Lambda}
\def\lb{\lambda}
\def\g{\gamma}
\def\vp{\varphi}
\def\a{\alpha}
\def\b{\beta}
\def\g{\gamma}

\def\s{\sigma}
\def\Si{\Sigma}
\def\om{\omega}

\def\d{\delta}
\def\t{\theta}
\def\k{\kappa}
\def\ve{\varepsilon}

\def\ts{\times}
\def\ify{\infty}
\def\fl{\forall}
\def\op{\oplus}
\def\ra{\rightarrow}
\def\wt{\widetilde}
\def\ot{\otimes}
\def\part{\partial}
\def\lgl{\langle}
\def\rgl{\rangle}

\def\sm{\simeq}
\def\sbs{\subset}

\def\un{{\rm 1\mkern-4mu I}}

\def\up#1{\raise 1ex\hbox{\sevenrm#1}}

\catcode`\@=11
\def\displaylinesno #1{\displ@y\halign{
\hbox to\displaywidth{$\@lign\hfil\displaystyle##\hfil$}&
\llap{$##$}\crcr#1\crcr}}

\vglue 1cm

\centerline{\twelvebf A Universal Action Formula}

\vglue 1cm

\centerline{Ali H. Chamseddine\up{1,2} \quad and \quad
Alain Connes\up{2}}

\vglue 1cm

\noindent {1. Theoretische Physik,
ETH-H\"onggerberg, CH-8093 Z\"urich, Switzerland}

\noindent {2. I.H.E.S., F-91440
Bures-sur-Yvette, France}

\vglue 3cm

\noindent {\parindent=1cm\narrower}
{\centerline {\bf Abstract.}} 
A universal formula for an action associated with
a noncommutative geometry, defined by a spectal triple  
$(\Ac ,\Hc ,D)$, is proposed.  It is based on the 
spectrum of the Dirac operator and is a geometric
invariant. The new symmetry principle is the 
automorphism of the algebra $\Ac $ which combines
both diffeomorphisms and internal symmetries. 
Applying this to the geometry defined by the
spectrum of the standard model gives an action
that unifies gravity with the standard model at
a very high energy scale.

\vglue 1cm
{\noindent PACS numbers: 02.40, 04.62, 12.10.}

\vfill\eject

Riemannian geometry has played an important role in
our understanding of space-time especially in the
development of the general theory of relativiy.
The basic data of Riemannian geometry consists in a {\it
manifold} $M$ whose points $x\in M$ are locally labelled by
finitely many coordinates $x^{\mu} \in \Rb$, and in the
infinitesimal {\it line element}, $ds^2 = g_{\mu \nu} \, 
dx^{\mu} \, dx^{\nu} \, .$
The dynamics of the metric is governed by the Einstein
action and the symmetry principle is diffeomorphism
invariance. The other basic interactions consisting
of strong, weak and electromagnetic forces are well
described by the standard model action and the symmetry
principle is the gauged internal symmetry. Therefore
the natural group of invariance that governs the sum
of the Einstein and standard model actions is the
semi-direct product
of the group of local gauge transformations, $\Uc = C^{\ify}
(M,U(1) \ts SU(2) \ts SU(3))$ by the natural action of
$\Diff (M)$.
Another concept which is vital to the
construction of the standard model is spontaneous
symmetry breaking and Higgs fields but this has no
geometrical significance.

\smallskip

In a new development in mathematics, it has been
shown that Riemannian geometry could be replaced
with a more general formulation, noncommutative
geometry. The basic data of noncommutative geometry consists
of an involutive {\it algebra} $\Ac$ of operators in Hilbert
space $\Hc$ and of a selfadjoint unbounded operator $D$ in
$\Hc$ [1-6].The inverse $D^{-1}$ of $D$ plays the role of the
infinitesimal unit of length $ds$ of ordinary geometry.

\smallskip

There is no information lost in trading the original Riemannian
manifold $M$ for the corresponding spectral triple $(\Ac
,\Hc ,D)$ where  $\Ac =
C^{\ify} (M)$ is the algebra  of smooth functions on $M$, 
$\Hc = L^2 (M,S)$ the Hilbert space of $L^2$-spinors and
$D$ the Dirac operator of the Levi-Civita Spin connection.
More importantly one can characterize the spectral triples
$(\Ac ,\Hc ,D)$ which come from the above spinorial
construction by very simple axioms ([4]) which involve the
dimension $n$ of $M$. The parity of $n$ implies a $\Zb /2$
grading $\g$ of the Hilbert space $\Hc$ such that,
$$
\g = \g^* \ , \ \g^2 =1 \ , \ \g a = a \g \quad \fl \, a \in
\Ac \ , \ \g D =-D \g \, . 
$$
Moreover one keeps track of the {\it real structure} on
$\Hc$ as an antilinear isometry $J$ in $\Hc$ satisfying
simple relations
$$
J^2 = \ve \ , \ JD = \ve' DJ \ , \ J\g = \ve'' \g J \ ; \
\ve ,\ve' ,\ve'' \in \{ -1,1\} 
$$
where the value of $\ve ,\ve' ,\ve''$ is determined by $n$
modulo 8.
 The usual emphasis on the points $x\in M$ of a
geometric space is now replaced by the spectrum $\Si \sbs
\Rb$ of the operator $D$. Indeed, if one forgets about the
algebra $\Ac$ in the spectral triple $(\Ac ,\Hc ,D)$ but
retains only the operators $D$, $\g$ and $J$ acting in $\Hc$
one can characterize this data by the spectrum
$\Si$ of $D$ which is a discrete subset with multiplicity of
$\Rb$. In the even case $\Si = -\Si$. The existence of
Riemannian manifolds which are isospectral (i.e. have the
same $\Si$) but not isometric shows that the following
hypothesis is stronger than the usual diffeomorphism
invariance of the action of general relativity,
$$
\hbox{``The physical action  depends only on} \ \Si \, .\hbox{``} 
$$
The most natural candidate for an action that depends only
on the spectrum is 
$$
\Trace \chi \left( {D\over \Lb} \right) + \lgl \psi ,D\psi
\rgl \, . \eqno (1)
$$
where $\chi $ is a positive function. This form of the
action is dictated by the condition that it is additive
for disjoint union.
In the usual Riemannian case the group $\Diff (M)$ of
diffeomorphisms of $M$ is canonically isomorphic to the group
$\Aut (\Ac)$ of automorphisms of the algebra $\Ac = C^{\ify}
(M)$. To each $\vp \in \Diff (M)$ one associates the algebra
preserving map $\a_{\vp} : \Ac \ra \Ac$ given by
$$
\a_{\vp} (f) = f \circ \vp^{-1} \qquad \fl \, f \in C^{\ify}
(M) = \Ac \, . 
$$
In general the group $\Aut (\Ac)$ of automorphisms of the
involutive algebra $\Ac$ plays the role of the
diffeomorphisms of the noncommutative (or spectral for
short) geometry $(\Ac ,\Hc ,D)$. The first interesting new
feature of the general case is that the group $\Aut (\Ac)$
has a natural normal subgroup,
$
\Int (\Ac ) \sbs \Aut (\Ac) ,$
where an automorphism $\a$ is {\it inner} iff there exists a
unitary operator $u\in \Ac$, $(uu^* = u^* u=1)$ such that,
$\a (a) = uau^* \qquad \fl \, a \in \Ac \, .$
This leads us to the postulate that:{\it The symmetry principle
in noncommutative geometry is invariance under the group $\Aut (\Ac)$.}
\smallskip

We now apply these ideas to derive a noncommutative
geometric action unifying gravity with the standard
model. The algebra is taken to be 
$\Ac = C^{\ify} (M) \ot \Ac_F $
where the algebra $\Ac_F$ is {\it finite dimensional},
$\Ac_F = \Cb \op \Hb \op M_3 (\Cb)$
and $\Hb \sbs M_2 (\Cb)$ is the algebra of quaternions,
$\Hb = \left\{ \left( \matrix{\a &\b \cr -\bar{\b} &\bar{\a}
\cr }  \right) \ ; \ \a ,\b \in \Cb \right\} \, . $
 $\Ac$   is a tensor product
 which geometrically corresponds to a product space, an
instance of spectral geometry for $\Ac$ is given by the
product rule,
$$
\Hc = L^2 (M,S) \ot \Hc_F \ , \ D = {\part \!\!\! /}_M \ot 1
+ \g_5 \ot D_F \eqno (2)
$$
where $(\Hc_F ,D_F)$ is a spectral geometry on $\Ac_F$,
while both $L^2 (M,S)$ and the Dirac operator ${\part \!\!\!
/}_M$ on $M$ are as above.

\smallskip

\noindent  The group
$\Aut (\Ac)$ of diffeomorphisms falls in equivalence classes
under the normal subgroup $\Int (\Ac)$ of inner
automorphisms. In the same way the space of metrics has a
natural foliation into equivalence classes. The {\it
internal fluctuations} of a given metric are given by the
formula,
$$
D=D_0 + A + JAJ^{-1} \ , \ A = \Si \, a_i [D_0 ,b_i] \ , \ a_i
, b_i \in \Ac \ \hbox{and} \ A=A^* \, . \eqno (3)
$$
Thus starting from $(\Ac ,\Hc ,D_0)$ with obvious notations,
one leaves the representation of $\Ac$ in $\Hc$ untouched
and just perturbs the operator $D_0$ by (3) where $A$ is an
arbitrary self-adjoint operator in $\Hc$ of the form $A=\Si
\, a_i [D_0 ,b_i] \ ; \ a_i ,b_i \in \Ac$. For Riemannian geometry
these fluctuations are trivial.

The
hypothesis which we shall test in this letter is that there
exist an energy scale $\Lb$ in the range $10^{15} -
10^{19} \Gev$ at which we have a  geometric action given by (1).

 The quarks $Q$ and leptons $L$ are :$Q = \left( \matrix{
u_L \cr d_L \cr d_R \cr u_R \cr
} \right) \ \  , L = \left( \matrix{
\nu_L \cr e_L \cr e_R \cr
} \right)$.

\noindent We now describe the internal geometry.
The choice of the Dirac operator and the action of
$\Ac_F$ in $\Hc_F$ comes from the restrictions that these must
satisfy:
$$
\displaylinesno{
J^2 = 1 \ , \ [J ,D] = 0 \ , \ [a,Jb^* J^{-1}] = 0 \cr
[[D,a] , J b^* J^{-1}]=0 \quad \fl \, a,b \, . &(4)
}
$$
We can now  compute the inner fluctuations of the
metric and thus  operators of the form: $A = \Si \, a_i
[D,b_i]$. This with the self-adjointness condition $A=A^*$
gives a $U(1)$, $SU(2)$ and $U(3)$ gauge fields as well as a
Higgs field. The computation of $A+ JAJ^{-1}$ removes a
$U(1)$ part from the above gauge fields (such that the full
matrix is traceless). The Dirac
operator $D_q$ that takes the inner fluctuations into account
is given by the $36 \ts 36$ matrix (acting on the 36 quarks)
(tensored with Clifford algebras)
$$
\eqalignno{
D_q & = \cr
&\left[ \matrix{
\g^{\mu} \ot \left(D_{\mu} \ot 1_2 -{i\over 2} g_{02}
A_{\mu}^{\a} \s^{\a} -{i \over 6} g_{01} B_{\mu} \ot 1_2
\right) \ot 1_3 , \ \g_5 \ot k_0^d \ot H , \ \g_5 \ot k_0^u
\ot \wt H \cr
\g_5 \ot k_0^{d*} \ot H^* , \qquad \qquad \g^{\mu} \ot
\left( D_{\mu} + {i \over 3} g_{01} B_{\mu} \right) \ot 1_3 ,
\qquad \qquad \qquad \qquad 0 \hfill \cr 
\g_5 k_0^{u*} \wt{H}^* , \qquad \qquad \qquad \qquad \qquad
0 , \qquad \qquad \g^{\mu} \ot \left( D_{\mu} -{2i \over 3}
g_{01} B_{\mu} \right) \ot 1_3 \hfill \cr } \right] \ot 1_3
\cr 
& + \g^{\mu} \ot 1_4 \ot 1_3 \ot \left( -{i \over 2} \,
g_{03} \, V_{\mu}^i \, \lb^i \right) &(5) \cr
}
$$
where $\s^{\a}$ are Pauli matrices and $\lb^i$ are
Gell-Mann matrices satisfying
$\Tr (\lb^i \lb^j) = 2\d^{ij} \, . $
The vector fields $B_{\mu}$, $A_{\mu}^{\a}$ and
$V_{\mu}^i$ are the $U(1)$, $SU(2)_w$ and $SU(3)_c$ gauge
fields with gauge couplings $g_{01}$, $g_{02}$ and $g_{03}$.
The differential operator $D_{\mu}$ is given by
$D_{\mu} = \part_{\mu} + \om_{\mu} $
where $\omega_{\mu}={1\over 4}\omega_{\mu}^{\ ab}\gamma_{ab}$
and $\g^{\mu} = e_a^{\mu} \, \g^a$. The scalar field $H$ is
the Higgs doublet, and $ \wt H = (i \s^2 H) \, $ is the $SU(2)$ conjugate
of $H$.

The Dirac operator acting on the leptons, taking inner
fluctuations into account is given by the  $9\ts 9$ matrix
(tensored with Clifford algebra matrices): 
$$
D_{\ell} = \left[ \matrix{
\g^{\mu} \ot \left( D_{\mu} -{i \over 2} g_{02} A_{\mu}^{\a}
\s^{\a} + {i \over 2} g_{01} B_{\mu} \ot 1_2 \right) \ot 1_3
\qquad \qquad \g_5 \ot k_0^e \ot H \cr
\g_5 \ot k_0^{*e} \ot H^* \hfill \qquad \qquad \qquad
\qquad \qquad \g^{\mu} \ot (D_{\mu} + ig_{01} B_{\mu}) \ot
1_3 \cr 
} \right] \, . \eqno (6)
$$
The matrices $k_0^d$, $k_0^u$ and $k_0^e$ are $3\ts 3$ family
mixing matrices.
According to our universal formula  the spectral
action for the standard model is given by:
$$
\Tr [ \chi (D^2 / m_0^2 )] + (\psi ,D \psi) \eqno
(7) 
$$
where $(\psi ,D\psi)$ will include both the quark 
and  leptonic interactions.
\noindent It is a simple exercise to compute the square
of the Dirac operator given by (5) and (6). This can be
cast into the elliptic operator form [7]:
$$
P=D^2 =-(g^{\mu \nu} \, \part_{\mu} \, \part_{\nu} \cdot
\un + \Ab^{\mu} \, \part_{\mu} + \Bb) 
$$
where $\un$, $\Ab^{\mu}$ and $\Bb$ are matrices of the
same dimensions as $D$.
Using the heat kernel expansion for
$$
\Tr e^{-tP} \sm \sum_{n\geq 0} t^{{n-m \over d}} \int_M
a_n (x,P) \, dv (x) 
$$
where $m$ is the dimension of the manifold in $C^{\ify}
(M)$, $d$ is the order of $P$ (in our case $m=4$,
$d=2$) and $dv (x) = \sqrt g \, d^m \, x$ where $g^{\mu
\nu}$ is the metric on $M$ appearing in $P$,
we can show that

$$
\Tr \chi (P) \sm \sum_{n\geq 0} f_n \, a_n (P) 
$$
where the coefficients $f_n$ are given by
$$
\displaylinesno{
f_0 = \int_0^{\ify} \chi (u) \, udu \ , \ f_2 =
\int_0^{\ify} \chi (u) \, du \ , \cr
f_{2(n+2)} = (-1)^n \, \chi^{(n)} (0) \ , \ n\geq 0
\cr }
$$
and $a_n (P) = \int a_n (x,P) \, dv (x)$ are 
the Seeley-de Witt coefficients. $a_n (P)$ vanish for odd
values of $n$ and the first four $a_n$ for even n are given
in Gilkey[7].
A very lengthy but straightforward calculation, the details
of which will be reported somewhere else [8] gives for the bosonic
action
$$
\eqalignno{
I_b = & \ \int d^4 x \, \sqrt g \ \biggl[ {1\over 2\kappa_0^2} \,
R - \mu_0^2 (H^* H) + a_0 \, C_{\mu \nu \rho \s} \, C^{\mu
\nu \rho \s} \cr
+ & \ b_0 \, R^2 + c_0 \, {}^* R {}^* R + d_0 \, 
R;_{\mu} {}^{\mu} \cr
+ & \ e_0 + {1\over 4} \, G_{\mu \nu}^i \, G^{\mu \nu i} +
{1\over 4} \, F_{\mu \nu}^{\a} \, F^{\mu \nu \a} \cr
+ & \ {1\over 4} \, B_{\mu \nu} \, B^{\mu \nu} + \vert
D_{\mu} \, H\vert^2 - \xi_0 \, R \vert H \vert^2 + \lb_0
(H^* H)^2 \biggl]+O\bigl({1\over{m_0^2}}\bigl) &(8) \cr
}
$$
where $C_{\mu \nu \rho \s}$ is the Weyl tensor and ${}^* R {}^* R$
is the Euler characteristic, and
$$
\eqalignno{
\mu_0^2 = & \ {4\over 3\kappa_0^2} \qquad
a_0 = \ -{9 \over 8g_{03}^2} \qquad
b_0 =  \ 0 &(9) \cr
c_0 = & \ -{11 \over 18} \, a_0 \qquad
d_0 =  \ -{2\over 3} \, a_0 \qquad
e_0 =  \ {45 \over 4\pi^2} \, f_0 \, m_0^4 \cr
\lb_0 = & \ {4\over 3} \, g_{03}^2 \, {z^2 \over y^4} 
 \qquad \xi_0 = \ {1\over 6} \, . \cr
}
$$
We have also denoted
$$
\eqalignno{
y^2 = & \Tr \left( \vert k_0^d \vert^2 + \vert k_0^u \vert^2
+ {1\over 3} \vert k_0^e \vert^2 \right) \cr
z^2 = & \Tr \left( (\vert k_0^d \vert^2 + \vert k_0^u
\vert^2)^2 + {1\over 3} \vert k_0^e \vert^4 \right) 
\cr
D_{\mu} \, H = & \, \part_{\mu} \, H - {i \over 2} \, g_{02}
\, A_{\mu}^{\a} \, \s^{\a} H - {i \over 2} \, g_{01}
\, B_{\mu} \, H \, . \cr
}
$$
The Einstein, Yang-Mills and Higgs term are normalized
 by taking:
$$
\eqalignno{
{15m_0^2 \, f_2 \over 4\pi^2} = & \ {1\over \kappa_0^2} \qquad
{g_{03}^2 \, f_4 \over \pi^2} = \ 1 \qquad
g_{03}^2 = \ g_{02}^2 = {5\over 3} \, g_{01}^2 \, . &(10)\cr
}
$$
Relations (10) among the gauge coupling constants coincide
with those coming from $SU(5)$ unification.

\smallskip

We shall adopt Wilson's view point of the renormalization
group approach to field theory [9] where the spectral action
is taken to give the {\it bare} action with bare quantities
$a_0, b_0, c_0 \ldots $  and at a cutoff scale $\Lb$ which regularizes
 the action the
theory is assumed to take a geometrical form. The
perturbative expansion is then reexpressed in terms of {\it
renormalized} physical quanties. The fields also receive
wave function renormalization.
\smallskip

\noindent The renormalized action receives counterterms of the same
form as the bare action but with physical parameters.
$a,b,c, \ldots .$
The renormalization group equations will yield
relations between the bare quantities and the physical
quantities with the addition of the cutoff scale $\Lb$.
Conditions on the bare quantities would translate into
conditions on the physical quantities. 
 The renormalization group equations of this system
were studied by Fradkin and Tseytlin [10] and is known to be
renormalizable, but non-unitary [11] due to the presence of
spin-two ghost (tachyon) pole near the Planck mass. We shall
not worry about non-unitarity (see, however, reference 12),
because in our view at the Planck energy the manifold
structure of space-time will break down and 
must be replaced with a genuienly noncommutative structure.

\smallskip

\noindent Relations between the bare gauge coupling
constants as well as equations (3.19) have to be imposed as
boundary conditions on the renormalization group equations
[9]. The bare mass of the Higgs field is related to the
bare value of Newton's constant, and both have quadratic
divergences in the limit of infinite cutoff $\Lb$. The
relations between $m_0^2, e_0$ and the physical quantities are:
$$\eqalignno{
m_0^2 &= m^2 \left( 1 + {\left( {\Lb_2 \over m^2} -1 \right)
\over 32\pi^2} \ \left( {9\over 4} \, g_2^2 + {3\over 4} \,
g_1^2 + 6\lb - 6k_t^2 \right)\right) + 0 \left( \ln \, {\Lb^2
\over m^2} \right) + \ldots \cr
e_0 &= e + {\Lb^4 \over 32\pi^2} \, (62) + \ldots &(11)\cr}
$$
For $m^2 (\Lb)$ to be small at low-energies $m_0^2$ should
be tuned to be proportional to the cutoff scale according to
equation (11).
Similarly the bare cosmological constant is
related to the physical one (which must be tuned to zero at
low energies). There is also a relation between the bare scale
$\k_0^{-2}$ and the physical one $\k^{-2}$ which is similar to
equation (11) (but with all one-loop contributions coming
with the same sign) which shows that $\k_0^{-1} \sim m_0$ and
$\Lb$ are of the same order as the Planck mass.

\smallskip

\noindent The renormalization group equations for the gauge
coupling constants $g_1, g_2, g_3$ are the same as those  
with $SU(5)$ boundary conditiions, and can be easily solved  
using the present experimental values for $\a_{em}^{-1} (M_Z)$
and $\a_3 (M_Z)$ to give [13]:
$$
\displaylinesno{
 \Lb  \sm  10^{15}(\Gev) \qquad 
 \sin^2 \t_w \sm 0.210 \, . &(12) \cr
}
$$
There is one further relation in our theory between the
$\lambda_0 (H^* H)^2$ coupling and the gauge couplings in equation (9)
imposed at the scale $\Lb$.
This relation could be simplified if we assume that the
top quark Yukawa coupling is much larger than all the
other Yukawa couplings. In this case the equation 
simplifies to
$\lb (\Lb) = {16\pi \over 3} \a_3 (\Lb) \, . $
Therefore the value of $\lb$ at the
unification scale is $\lb_0 \sm 0.402$ showing that one
does not go outside the perturbation domain. In reality  the RG
equations for $\lb$ and $k_t$ togother with the boundary condition
on $\lambda $ could be used to determine the Higgs mass
at the low-energy scale $M_Z$. The renormalization
group equations are complicated and must be integrated
numerically [14]. One can read the solution for the Higgs mass
from the study of the triviality bound and this gives
$m_H =170-180$ Gev. One expects this prediction to be
correct to the same order as that of $\sin^2 \t_w$ which
is off from the experimental value by ten percent.
Therefore the bare action we obtained and associated with
the spectrum of the standard model is consistent 
within ten percent provided the cutoff scale is taken to be
$\Lb \sim 10^{15} \Gev$ at which the action becomes 
geometrical.
  
 There is, however, a stronger disagreement where
Newton's constant comes out to be too large. This is so
because the gravity
sector requires the cutoff scale to be of the same order as
the Planck scale while the condition on gauge coupling
constants give $\Lb \sim 10^{15} \Gev$. One easy way
to avoid this problem is to assume that the spectrum
contains in addition a fermionic particle which only
interacts gravitationally (such as a gravitino), but
at present we shall not commit ourselves.
Incidently the porblem
 that Newton's constant is 
coming out to be too large is also present  in string
theroy where  also has unification of gauge couplings and
Newton's constant occurs [15].  These results must be taken as an
indication that the spectrum of the standard model has to be
altered as we climb up in energy. The change may happen at
low energies (just as in supersymmetry which also pushes the
cutoff scale to $10^{16} \Gev$) or at some intermediate
scale. 
Ultimately one would hope that modification of the
spectrum will increase the cutoff scale nearer to the Planck
mass as dictated by gravity.
\medskip

To summarize, we have shown that the basic symmetry for a noncommutative
space $(\Ac ,\Hc ,D)$ is $\Aut (\Ac)$. This symmetry
includes diffeomorphisms and internal symmetry
transformations. The bosonic action is a spectral
function of the Dirac operator while the
fermionic action takes the simple linear form $(\psi
,D\psi)$ where $\psi$ are spinors defined on the Hilbert
space. Applying this principle to the  noncommutative
geometry of the standard model gives the standard model
action coupled to Einstein and Weyl gravity plus higher order
non-renormalizable interactions supressed by powers of the
inverse of the mass scale in the theory.
There are some relations between the
bare quantities. The renormalized action will have the same
form as the bare action but with physical quantities
replacing the bare ones.  The relations among the bare
 quantities must be
taken as boundary conditions on the renormalization group
equations governing the scale dependence of the physical
quantities. These boundary condition imply that the
cutoff scale is of order $\sim 10^{15} \Gev$ and 
 $\sin^2 \t_w \sim 0.21$
which is off by ten percent from the true value. We also have
a prediction of the Higgs mass in the interval $170-180 
\Gev$. This can be taken as an indication that the
noncommutative structure of space-time reveals itself at
such high scale where the effective action has a geometrical
interpretation.
The slight disagreement with experiment
indicates that the spectrum of the standard model could
not be extrapolated to very high energies without adding
new particles necessary to change the RG equations of the
gauge couplings. One possibility could be supersymmetry,
but there could be also less drastic solutions. This also
could be taken as an indication that the the concept of
space-time as a manifold breaks down and the
noncommutativity of the algebra must be extended to include
the manifold part.  Finally, we hope that our
universal action formula should be applicable to many
physical systems of which the most important could be
superconformal field theory where the supersymmetry
charge play the role of the Dirac operator. 

\vglue 1cm

\noindent {\bf Acknowledgments.} A.H.C. would like to
thank J\"urg Fr\"ohlich for very useful discussions and
I.H.E.S. for hospitality where part of this work was done.

%\vfill\eject
\vglue 1cm

\noindent {\bf References.}

\medskip

\item{[1]}  A. Connes, {\it Publ. Math. IHES} {\bf 62},
44 (1983); Noncommutative Geometry (Academic Press,
New York 1994).

\item{[2]} A. Connes and J. Lott, {\it Nucl. Phys. Proc.
Supp.} {\bf B18}, 295 (1990); proceedings of 1991
Carg\`ese Summer Conference, edited by J. Fr\"ohlich et
al (Plenum, New York 1992).

\item{[3]} D. Kastler, {\it Rev. Math. Phys.} {\bf 5}, 477
(1993).

\item{[4]} A. Connes, Gravity coupled with matter and the
foundation of noncommutative geometry, hep-th/9603053.

\item{[5]} A.H. Chamseddine, G. Felder and J. Fr\"ohlich,
{\it Comm. Math. Phys.} {\bf 155}, 109 (1993); A.H.
Chamseddine, J. Fr\"ohlich and O. Grandjean, {\it J.
Math. Phys.} {\bf 36}, 6255 (1995).

\item{[6]} A. Connes, {\it J. Math. Phys.} {\bf 36}, 6194
(1995).

\item{[7]} P. Gilkey, Invariance theory, the heat
equation and the Atiyah-Singer index theorem, (Publish or
Perish, Dilmington, 1984).

\item{[8]} A. H. Chamseddine and A. Connes, "The Spectral
Action Principle", hep-th/9606001.

\item{[9]} K.G. Wilson, {\it Rev. Mod. Phys.} {\bf 47}
(1975), 773; For an exposition very close to the steps taken
here see C. Itzykson and Drouffe, {\it Field theory},
Chapter five, Cambridge University Press.

\item{[10]} E. Fradkin and A. Tseytlin, {\it Nucl. Phys.}

\item{[11]} K.S. Stelle, {\it Phys. Rev.} {\bf D16}, 953
(1977).

\item{[12]} E. Tomboulis, {\it Phys. Lett.} {\bf 70B},
361 (1977).

\item{[13]} For a review see G. Ross, Grand unified
theories, {\it Frontiers in Physics Series}, Vol.60
(Benjamin, New York).

\item{[15]} M. B\'eg, C. Panagiotakopoulos and A. Sirlin,
{\it Phys. Rev. Lett.} {\bf 52}, 883 (1984); M. Lindner,
{\it Z. Phys.} {\bf C31}, 295 (1986).

\item{[14]} E. Witten, "Strong Coupling Expansion of Calabi-Yau
Compactification", hep-th/9602070.

\bye